\documentclass[conference]{IEEEtran}
\IEEEoverridecommandlockouts

\usepackage{cite}
\usepackage{amsmath,amssymb,amsfonts}
\usepackage{algorithmic}
\usepackage{graphicx}
\usepackage{textcomp}
\usepackage{xcolor}
\usepackage{svg}
\usepackage{soul}
\usepackage{tabularx}
\usepackage{url}
\usepackage{siunitx}
\usepackage[section]{placeins}
\usepackage{filecontents}
\usepackage{enumitem}
\usepackage[caption=false,font=footnotesize]{subfig}

\newcommand{\system}{{\textsc{Hilo}\space}}
\newcommand{\systemnospace}{\textsc{Hilo}}
\newcommand{\fakeparagraph}[1]{\vspace{.5mm}\noindent\textbf{#1.}}
\newcommand{\fakepar}[1]{\fakeparagraph{#1}}
\def\BibTeX{{\rm B\kern-.05em{\sc i\kern-.025em b}\kern-.08em
    T\kern-.1667em\lower.7ex\hbox{E}\kern-.125emX}}

\title{\systemnospace: Enabling Low-power, Dual-Band Communication using Tunnel Diode Oscillators}

\author{\IEEEauthorblockN{Dhairya Shah\textsuperscript{\textsection},
Rajashekar Reddy Chinthalapani\textsuperscript{\textsection},
Pramuka Sooriya Patabandige, 
Ambuj Varshney}
\IEEEauthorblockA{
National University of Singapore, Singapore\space
\\
Email: \{dhairya, rajashekar.c, pramuka\}@u.nus.edu, ambujv@nus.edu.sg}}

\begin{document}
\maketitle

\begingroup\renewcommand\thefootnote{\textsection}
\footnotetext{Co-primary authors with equal contributions to this work.}

\begin{abstract}
Wireless communication remains the most power-consuming operation in embedded systems. Low-power transmitters such as backscatter achieve microwatt-scale operation but produce weak signals susceptible to in-band interference and frequency-selective fading, particularly in non-line-of-sight settings. Multi-band transmission can mitigate these effects through spectral-domain redundancy rather than temporal-domain redundancy, improving reliability without sacrificing bitrate or latency. However, conventional multi-band radio transceivers require duplicated RF chains or power-hungry synthesizers. We introduce \systemnospace, a tunnel diode-based frontend that enables dual-band operation from a single circuit. \system inverts the conventional paradigm: rather than suppressing the harmonics inherent to tunnel diode nonlinearity, it harnesses them. An external device injection-locks the oscillator to a fundamental frequency, simultaneously stabilizing a higher harmonic and yielding two injection-locked carriers without a second oscillator or frequency multiplier. In transmit mode, \system exploits self-oscillating mixing to modulate both carriers simultaneously; in receive mode, it uses autodyne downconversion at both frequencies. By turning an inherent nonlinearity into a feature, \system achieves dual-band links while consuming under \SI{210}{\micro\watt}.

\end{abstract}

\section{Introduction}
Wireless communication is the most power-intensive operation in embedded systems. Despite sustained efforts in designing low-power transmitters, reducing energy consumption comes at the cost of link margin, making communication reliability a major challenge~\cite{ambientbackscatter,lorea,backscatter2006}. This is especially true for transmitters based on backscatter mechanisms~\cite{ensworth2015every} and tunnel diodes~\cite{varshney2022judo,varshney2019tunnelscatter}, whose radiated signal strength is weaker than that of nearby active radio transmitters, making them susceptible to interference. The challenge is particularly severe in non-line-of-sight~(NLoS) environments, where frequency-selective fading can create deep nulls, forcing conservative bitrates and retransmissions. For low-power communication, \textit{reliability is a prerequisite for energy efficiency}.

Improving the reliability of weak-signal transmissions typically involves tradeoffs. Coding schemes sacrifice effective bitrate by adding payload redundancy~\cite{varshney2017towards}. Advanced post-processing such as machine learning increases receiver complexity and power consumption. \system proposes a different approach: multi-band communication. By simultaneously transmitting and receiving replica information across multiple frequency bands, \system introduces redundancy in the spectral domain rather than the temporal domain. This mitigates narrowband interference and reduces the probability that frequency-selective fading degrades all channels simultaneously. Because the bands operate in parallel, link reliability improves  without incurring latency or bitrate penalties.

\begin{figure}[!t]
    \centering
    \subfloat[\system frontend as a transmitter]{%
        \includegraphics[width=\linewidth]{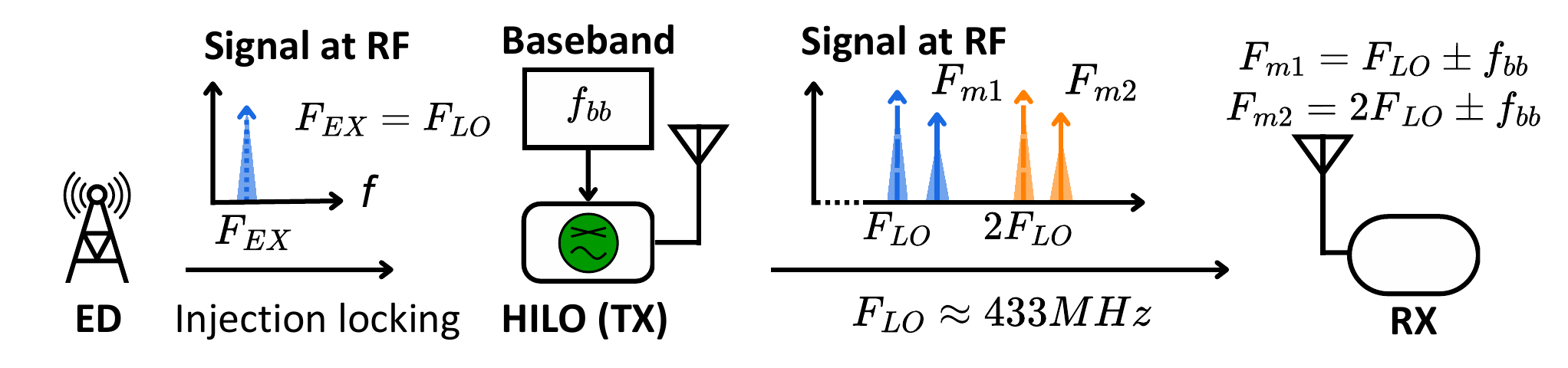}%
        
        \label{fig:HILO-TX}
    }
    \vspace{-0.7mm}
    \subfloat[\system frontend as a receiver]{%
        \includegraphics[width=\linewidth]{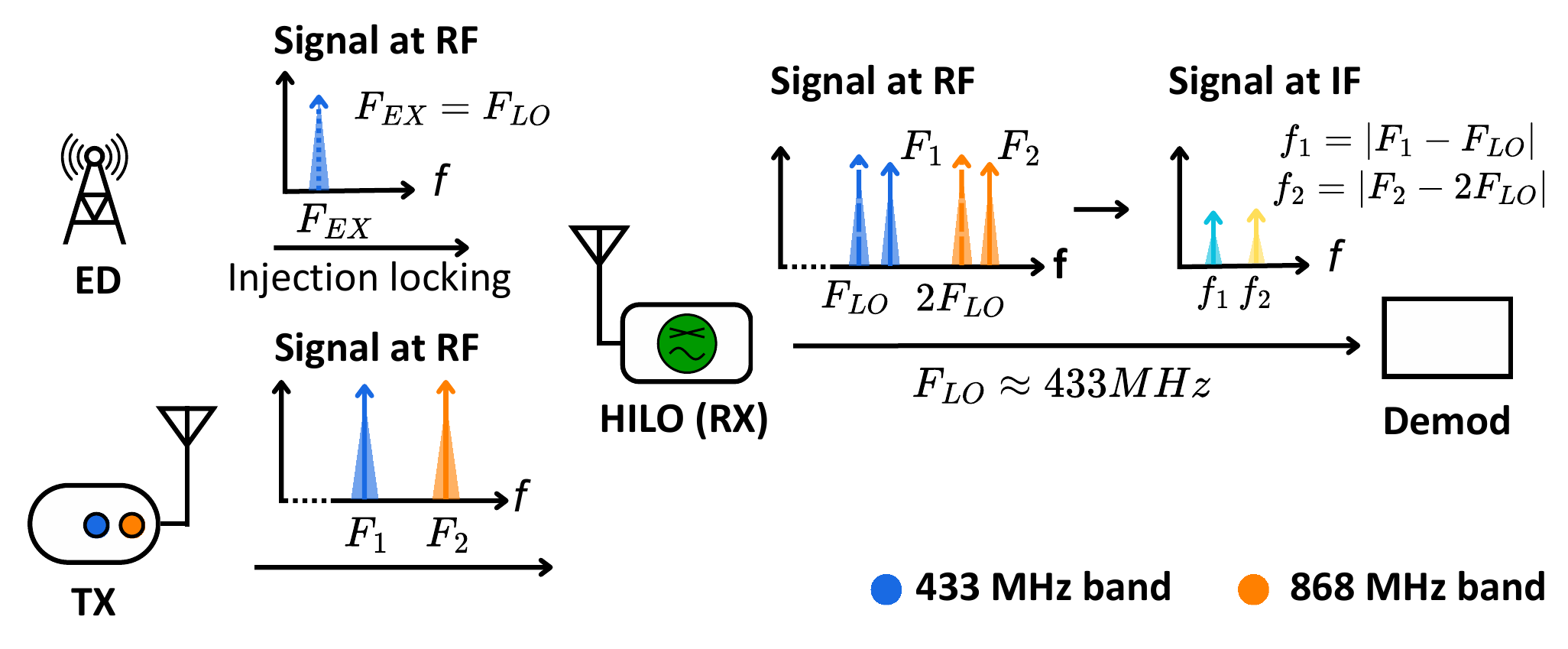}%
        
        \label{fig:HILO-RX}
    }
    \vspace{-2mm}
   \caption{   A carrier emitter device injection-locks the tunnel diode oscillator at $F_{\mathrm{LO}}\approx\SI{433}{\mega\hertz}$, which also generates a harmonic at $2F_{\mathrm{LO}}\approx\SI{868}{\mega\hertz}$. (a) \system in transmit mode mixes a baseband $f_{\mathrm{bb}}$ signal, generating sidebands around the carrier and its harmonic for dual-band transmission. (b) \system in receive mode is sensitive to signals near $F_{\mathrm{LO}}$ and $2F_{\mathrm{LO}}$, using the autodyning property to downconvert signals in both bands.}

\vspace{-6mm}
    \label{fig:OVERVIEW}
\end{figure}

Generating signals across multiple bands is incompatible with the power budget of typical embedded devices. Conventional transmitter architectures rely on phase-locked loops, frequency multipliers, and complex control logic; scaling these to support multiple bands increases energy overhead and circuit complexity. Recent backscatter systems~\cite{Ye2023,HarmonicWisp,TagVibra} attempt to bypass this by exploiting tag nonlinearity to generate harmonics. However, these systems remain tethered to the delegation architecture, requiring a strong, proximate emitter device and suffering from the short transmission ranges intrinsic to conventional backscatter mechanism~\cite{varshney2019tunnelscatter}. Furthermore, these prior efforts also focuses exclusively on the uplink. Real-world systems require bidirectional communication, a capability that existing multi-band low-power systems lack.

This paper introduces \systemnospace, which stands for \ul{h}armonic \ul{i}njection-\ul{l}ocked \ul{o}scillator, a low-power system that enables dual-band transmission and downconversion on a microwatt power budget. \system achieves this by leveraging the inherent nonlinearity of tunnel diode. Recent tunnel diode oscillator~(TDO) systems~\cite{somix,varshney2022judo} generate both a fundamental carrier and its harmonics, but treat the harmonics as unintended and harmful emissions, employing filters and other techniques to suppress them. \system inverts this paradigm. Instead of discarding these harmonics, we harness them. We demonstrate that naturally generated harmonics can be used to establish frequency diversity without increasing circuit complexity.

The tunnel diode is a highly nonlinear device. When coupled with a resonant LC tank, it generates not only a fundamental frequency but also harmonic content. Recent systems that stabilize TDOs without injection locking, such as AudioCast~\cite{audiocast} and M\textsuperscript{2}~\cite{m2_mobisys26}, show visible harmonics in their  spectra. While SharpPeak~\cite{Padmal2025_thesis} reports a harmonic-free spectrum, the cubic-like I-V characteristic of the tunnel diode makes harmonic generation unavoidable in principle. Measurement conditions strongly influence whether harmonics are observed. SharpPeak's spectra are captured over the air rather than via a conducted measurement at the oscillator terminals, so the reported suppression reflects the combined response of the TDO, matching network, antenna, and free-space propagation, all of which attenuate energy at $2F_{LO}$ more than at $F_{LO}$. Although conventional designs treat these harmonics as noise to be filtered, \system exploits them to add spectral redundancy.

A free-running TDO is unstable and drifts with environmental changes~\cite{tunnelsense}. \system shows that injection-locking the fundamental simultaneously stabilizes the harmonics, yielding two frequency-locked carriers without a second oscillator or frequency multiplier. This allows the emitter device to operate at the lower frequency, benefiting from superior propagation while still stabilizing the high-frequency band.

\system builds on recent work demonstrating the self-oscillating mixing~(SoM) property of TDOs~\cite{varshney2022judo}. \system shows that SoM modulates not only the fundamental but also the harmonic, enabling  dual-band modulation without incurring any additional energy cost. Building on this propery, \system modulates the carrier from TDO using frequency-shift keying (FSK) and demonstrate  communication over tens of meters.

Finally, reception is essential for a complete communication system. Recent work demonstrates the ability of tunnel diodes to downconvert transmissions to intermediate frequency for  receiver design~\cite{somix}. \system extends this capability, showing that harmonics generated by the TDO also enable downconversion via autodyning, paving the way for dual-band reception. Together, these contributions yield a dual-band tunnel-diode based transceiver frontend  while consuming only \SI{210}{\micro\watt}.

\section{Related Work}
\label{sec:background}

\fakepar{Harmonic backscatter}
Recent work exploits the inherent nonlinearity in backscatter mechanisms. In RFID systems, a reader interrogates the tag with a fundamental tone, and the tag backscatters a response at a harmonic, typically the second~\cite{HarmonicWisp,Ye2023,TagVibra}. This frequency separation mitigates the strong self-interference that plagues conventional backscatter links and improves robustness in cluttered environments~\cite{HarmonicWisp}. Although harmonics were historically treated as parasitic, recent systems intentionally leverage harmonic backscatter for detection, localization, and sensing~\cite{TagVibra,passivevibration,harmonictunnel}. TagVibra~\cite{TagVibra} extracts phase from harmonic backscatter for sub-millimeter vibration sensing, Palazzi et al.~\cite{passivevibration} use amplitude-modulated harmonic backscatter for structural monitoring, and Piumwardane et al.~\cite{dilushi_harmonic_sensing} demonstrate high-resolution analog backscatter sensing. These systems are passive, using harmonics primarily for carrier self-interference isolation or sensing fidelity rather than to improve link reliability. Moreover, they inherit the limitations of backscatter: proximity to the emitter device and unidirectional communication (tag-to-reader only). \system builds on these insights while addressing both limitations. It introduces a tunnel diode-based  frontend that  exploits TDO nonlinearity for dual-band operation and supports bidirectional communication through autodyning and SoM.

\begin{figure}[!t]
    \centering
    \includegraphics[width=0.95\linewidth]{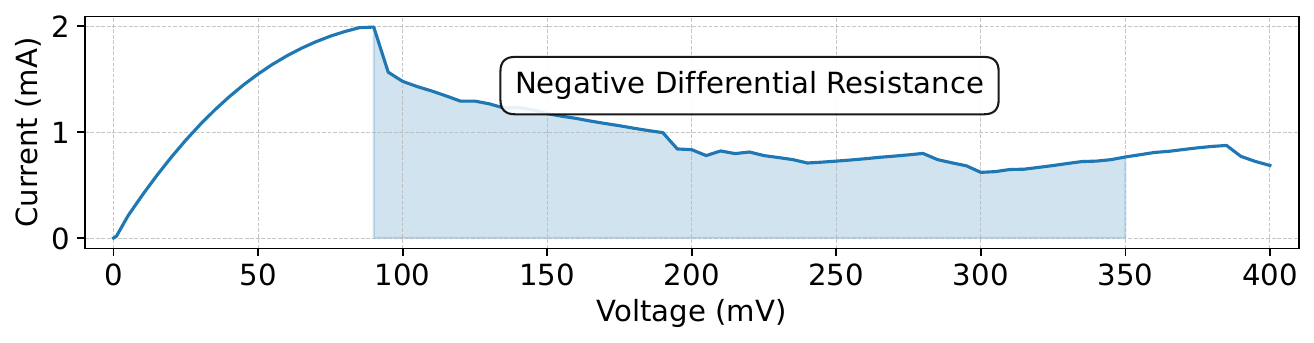}
    \vspace{-4mm}
    \caption{I-V characteristics of GaAs 3I306E tunnel diode. The tunnel diode is biased in the negative differential resistance region and coupled with an LC resonant tank circuit to generate oscillations.}
    \vspace{-6mm}
    \label{fig:ivcurve}
\end{figure}

\fakepar{Tunnel diodes}
They exhibit negative differential resistance~(NDR) in their I-V characteristics, as shown in Figure~\ref{fig:ivcurve}. Because NDR occurs at very low voltages and currents, tunnel diodes are attractive for low-power RF designs. Previous work has exploited this property primarily for backscatter amplification. Amato et al. demonstrated up to \SI{40}{\decibel} gain using tunnel diode-based reflection amplifiers in a \SI{5}{\giga\hertz} backscatter tag~\cite{amato2015long,amato2018tunneling}, while Adeyeye et al. designed a low-power repeater for RFID tags~\cite{adeyeye20215}. Eid et al. developed RFID tags that harvest energy from the carrier signal and use the tunnel diode for reflection amplification~\cite{eid}. Amato et al. further proposed the harmonic tunneling tag, which biases a tunnel diode in its nonlinear region to achieve second-harmonic conversion gain over conventional passive harmonic backscatter~\cite{harmonictunnel}. \system builds on these efforts but focuses on improving communication link reliability.

Recent efforts goes beyond backscatter, using tunnel diodes to build TDOs for transmitters~\cite{varshney2022judo,tunnelifi,audiocast} and receivers~\cite{somix}. These achieve microwatt-level operation but sacrifice frequency stability. Varshney et al. address this instability through injection locking~\cite{varshney2022judo}, while Thaddeus et al. exploit the frequency drift itself for vital-sign sensing~\cite{tunnelsense}. The same nonlinearity that causes instability also generates strong harmonic components, as oscillations deviate from a pure sinusoid~\cite{tunnelharmonic}. Previous designs suppress these harmonics for  purity; \system harnesses them to realize dual-band operation.

\section{Design}
\label{sec:design}
\system comprises three elements: a tag that generates carriers at both $F_{LO}$ (\SI{433}{\mega\hertz}) and $2F_{LO}$ (\SI{868}{\mega\hertz}), an emitter device~(ED) that injection-locks the tag at $F_{LO}$, and commodity receivers (e.g., TI CC1310) at one or both bands. In transmit mode, the TDO performs SoM with an input baseband signal, producing  sidebands around both $F_{LO}$ and $2F_{LO}$ for concurrent dual-band transmission. In receive mode, a conventional transmitter sends at either $F_{LO}$ or $2F_{LO}$, and the TDO downconverts the incoming signal to IF via autodyning for demodulation. When both bands are decodable, the receiver gains a frequency-diversity channel that improves link reliability. Figure~\ref{fig:OVERVIEW} provides the system overview.

\begin{figure}[!t]
    \centering
    \subfloat[Circuit Schematic]{
    \includegraphics[width=0.6\linewidth]{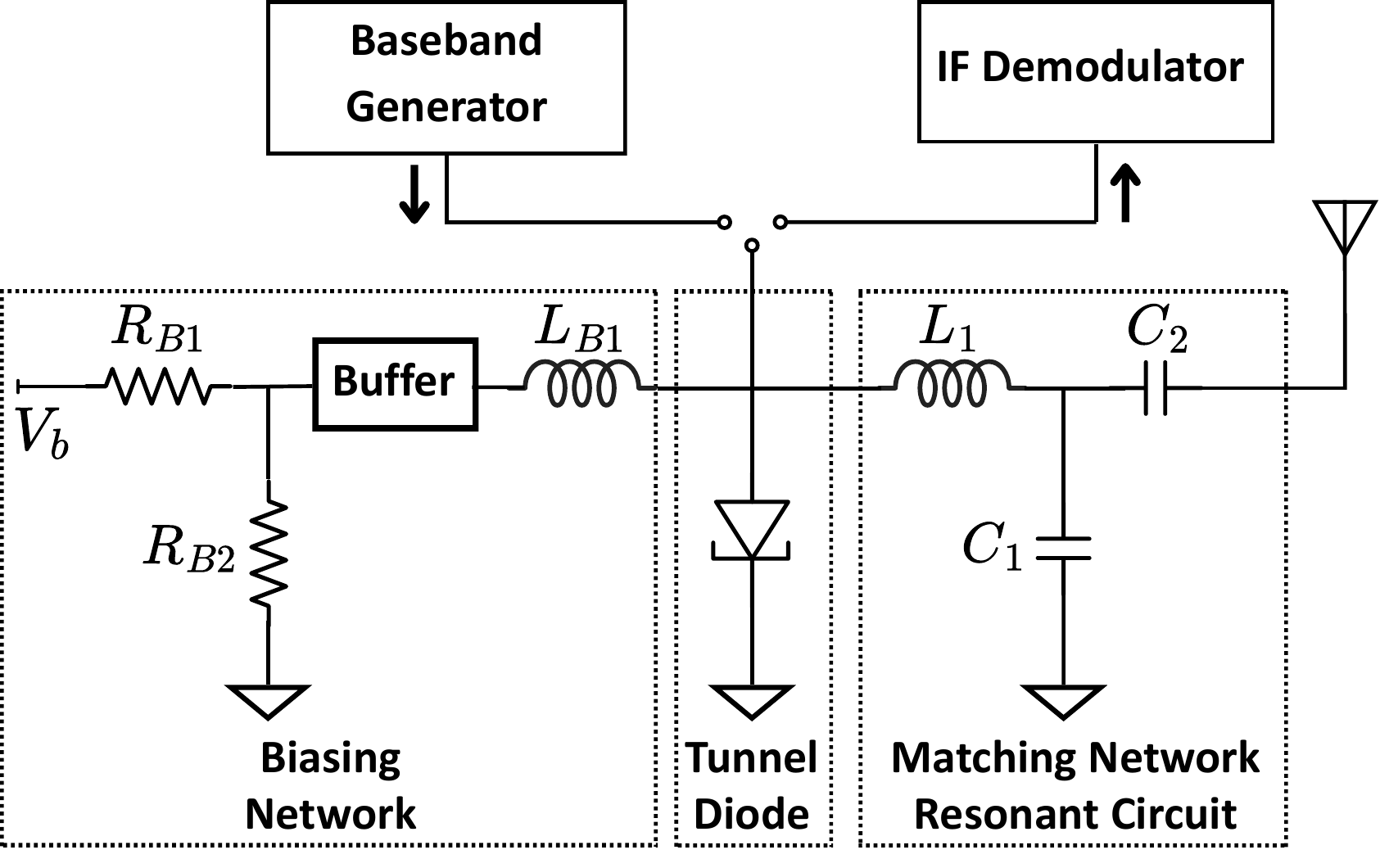}
    \label{fig:HILO-CIRCUIT}}
    \subfloat[Prototype]{
    \includegraphics[width=0.3\linewidth]{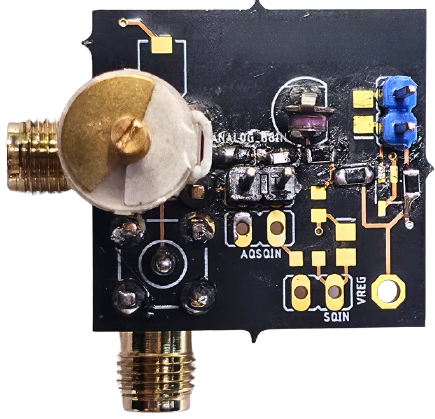}
    \label{fig:HILO-PCB}}
    \vspace{-2mm}
    \caption{The \system frontend consists of a tunnel diode coupled with an LC matching resonant tank, resulting in a TDO. Its nonlinear characteristics generate a fundamental carrier in the \SI{433}{\mega\hertz} band along with its second harmonic in the \SI{868}{\mega\hertz} band, enabling simultaneous dual-band operation.}
    \vspace{-4mm}
    \label{fig:HILO-ckt-proto}
\end{figure}

\subsection{Dual-band Carrier Generation}
\system is designed using a TDO tuned to the \SI{433}{\mega\hertz} ISM band by configuring the LC-tank circuit~(Figure~\ref{fig:HILO-CIRCUIT}). Because the tunnel diode is a strongly nonlinear device, its steady-state oscillation is not purely sinusoidal. The output spectrum contains a strong fundamental at $F_{LO}$ and higher-order components at integer multiples $kF_{LO}$. In our design, the second harmonic at $2F_{LO}$ falls naturally within the \SI{868}{\mega\hertz} ISM band, enabling dual-band carrier generation without an additional oscillator or mixing chain. We implement the \system frontend with commercial off-the-shelf components on an FR-4 PCB (Figure~\ref{fig:HILO-PCB}). The TDO uses a GE 3I306E tunnel diode~\cite{TunnelDiode3I306E}. Two SMA connectors couple directly to the tank circuit, allowing simultaneous connection of the ED and a spectrum analyzer during cabled characterization.

\begin{figure}[!t]
    \centering
    \includegraphics[width=\linewidth]{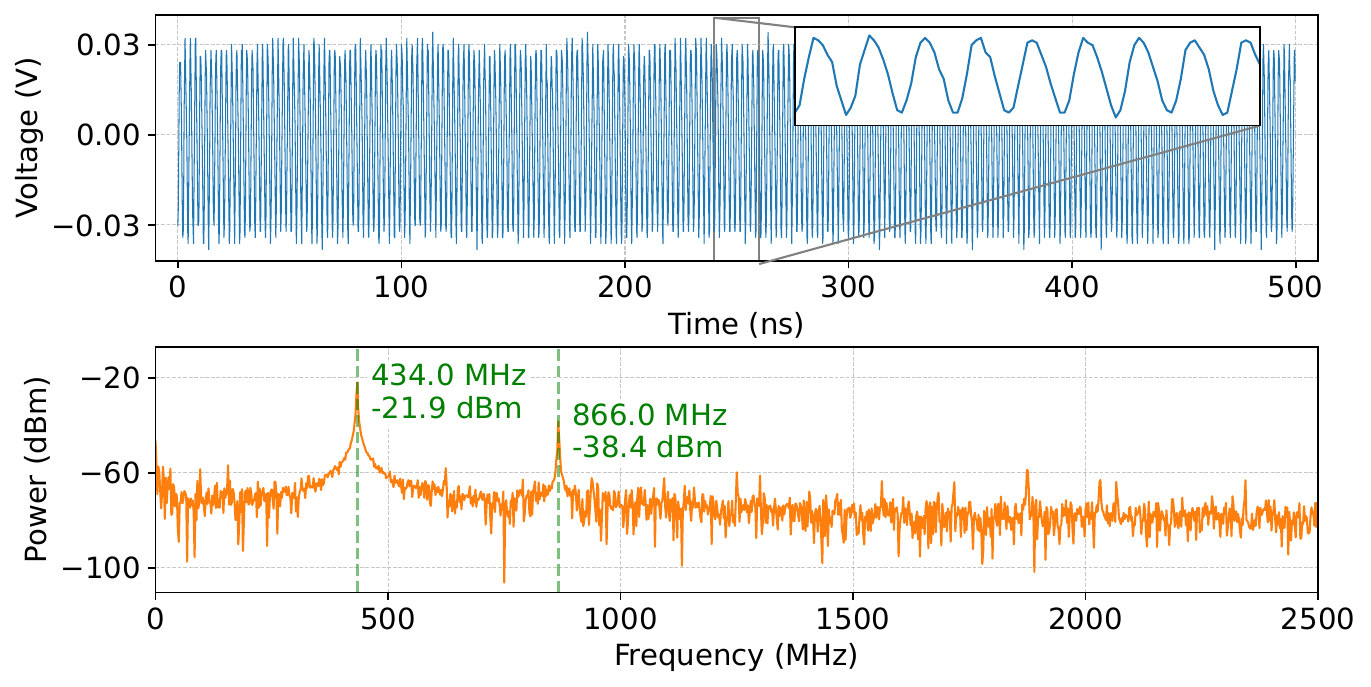}
    \vspace{-8mm}
    \caption{TDO tuned near \SI{433}{\mega\hertz} produces a harmonic in the \SI{868}{\mega\hertz} frequency band, resulting in a dual-band carrier source using single component.}
    \vspace{-6mm}
    \label{fig:timedomain}
\end{figure}
Figure~\ref{fig:timedomain} presents the time-domain representation of the output of the TDO and its corresponding  spectrum. The spectrum is computed via an FFT (length $N=2500$) of the waveform captured at \SI{5}{GSa/s} using a Keysight DSOX3104T oscilloscope with \SI{1}{\giga\hertz} bandwidth, connected via SMA cable with \SI{50}{\ohm} input termination. The resulting spectrum shows a fundamental tone near \SI{434}{\mega\hertz} and a harmonic at \SI{866}{\mega\hertz}, approximately \SI{16}{\decibel} below the fundamental frequency.

\fakepar{Understanding nonlinearity and harmonics} The generation of the useful second harmonic arises from the distinct $I$--$V$ characteristic of the tunnel diode. The current through the diode can be approximated by a polynomial expansion
$i(v) = a_1 v + a_2 v^2 + a_3 v^3 + \dots$ 
around the bias point~\cite{tunnelharmonic}. Higher-order terms in this expansion can also produce additional harmonic and intermodulation components; in this work, we focus on the dominant second-harmonic component that enables dual-band operation. When the tank circuit oscillates at the fundamental frequency $v(t) = A \cos(\omega_0 t)$, where $\omega_0 = 2\pi F_{LO}$, the quadratic term $a_2 v^2$ produces a component proportional to $\cos^2(\omega_0 t) = \frac{1}{2}(1 + \cos(2\omega_0 t))$. This $2\omega_0$ term is in the \SI{868}{\mega\hertz} band. While typically considered parasitic interference and suppressed, \system uses this component~(\SI{-38.4}{\decibel m}) to serve as both the carrier for transmission and the effective local oscillator for downconversion.

\begin{figure}[!t]
    \centering
    \includegraphics[width=0.95\linewidth]{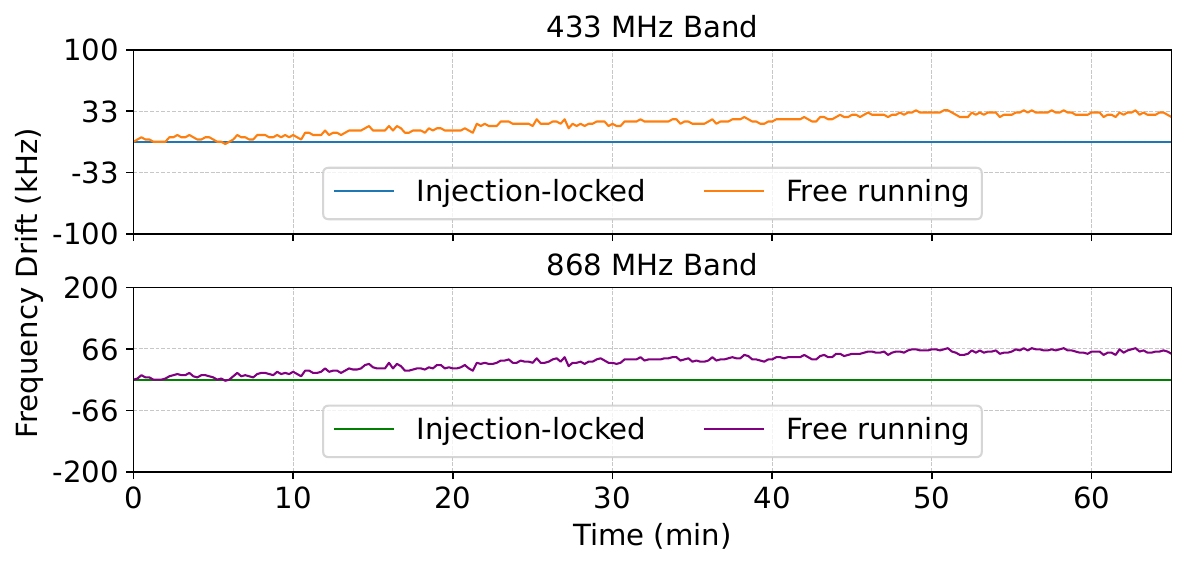}
    \vspace{-4mm}
    \caption{Frequency stability analysis showing the drift of the fundamental and harmonic bands over time. Injection locking at $F_{LO}$ stabilizes both the fundamental and the $2 F_{LO}$ harmonic using a single ED reference.}
    \vspace{-4mm}
    \label{fig:stability}
\end{figure}

\fakepar{Carrier frequency stability} A known challenge with TDOs is poor frequency stability~\cite{varshney2022judo,somix,audiocast,tunnelsense}. As shown in Figure~\ref{fig:stability}, free-running TDOs drift significantly over time due to sensitivity to bias conditions and environmental perturbations such as temperature and antenna loading. While recent efforts~\cite{audiocast,m2_mobisys26} have addressed this instability without external carriers, \system stabilizes the TDO using an external carrier signal from an ED to injection-lock the oscillator. In injection locking, the TDO synchronizes its phase and frequency to the injected carrier. The harmonic component inherits this stability because it is  generated from the locked fundamental. To quantify this, we apply a \SI{433}{\mega\hertz} injection tone using a USRP SDR and monitor the output with a signal analyzer. As shown in Figure~\ref{fig:stability}, injection locking at \SI{433}{\mega\hertz} stabilizes both the fundamental and the \SI{868}{\mega\hertz} harmonic, providing two frequency-locked carriers from a single ED.

\subsection{Upconversion for Transmission}
To enable transmission, \system leverages the SoM property of the TDO~\cite{varshney2022judo}. The diode's nonlinear $I$--$V$ characteristic multiplies the TDO voltage with the  baseband signal, producing  sidebands around both the fundamental and its harmonic.

\begin{figure}[!t]
    \centering
    \includegraphics[width=\linewidth]{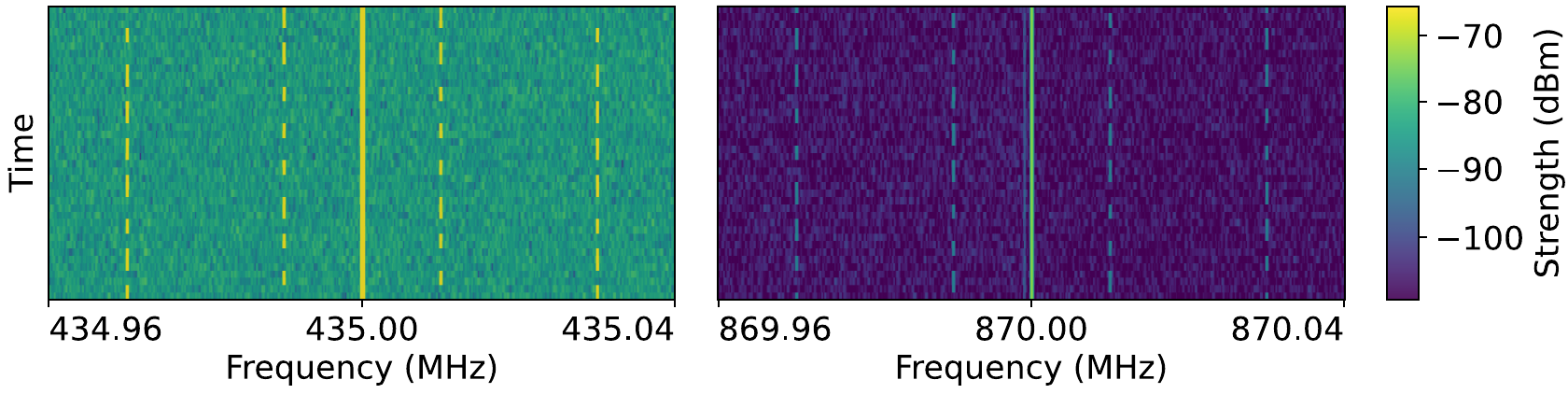}
    \vspace{-8mm}
    \caption{The same baseband modulation applied once is translated to both \SI{433}{\mega\hertz} and \SI{868}{\mega\hertz} bands using a single \system frontend.}
    \vspace{-6mm}
    \label{fig:dualband_waterfall}
\end{figure}

\fakepar{Simultaneous upconversion} The \system  simultaneously upconverts the baseband signal to both the fundamental ($F_{LO}$) and the harmonic ($2F_{LO}$). Unlike direct frequency modulation, where deviation scales with harmonic order, SoM acts as a mixer at both frequencies. The baseband frequency $f_{bb}$ is preserved in the translation, producing sidebands at $F_{m1} = F_{LO} \pm f_{bb}$ and $F_{m2} = 2F_{LO} \pm f_{bb}$. We verify this experimentally by injecting a  FSK baseband signal from an Analog Discovery 3 waveform generator at the \system input port. The injection-locking signal is provided via a USRP in a cabled setup, and a spectrum analyzer captures the output. The resulting spectrum (Figure~\ref{fig:dualband_waterfall}) confirms that the sideband offset from the carrier remains identical in both  bands.

\fakepar{Characterizing upconversion} We characterize upconversion efficiency using the same cabled setup described previously, with a USRP providing the injection-locking signal and a spectrum analyzer capturing the output. An Analog Discovery 3 provides a sinusoidal baseband input, swept in frequency and amplitude. Figure~\ref{fig:ss_vs_bb_freq} shows mixed signal strength against baseband frequency. The response remains relatively flat up to \SI{1}{\mega\hertz}, confirming sufficient bandwidth to support higher data rates without significant attenuation. Figure~\ref{fig:ss_vs_bias_voltage} shows the linearity of the mixing process. As the peak-to-peak voltage of the baseband signal increases, sideband power increases nearly linearly. Consistent with the harmonic generation process, upconverted signal strength at \SI{868}{\mega\hertz} is approximately \SI{10}{\decibel} lower than at the fundamental but remains well above the sensitivity threshold of standard receivers.

\begin{figure}[!t]
    \centering
    \subfloat[Baseband frequency response]{%
        \includegraphics[width=0.95\linewidth]{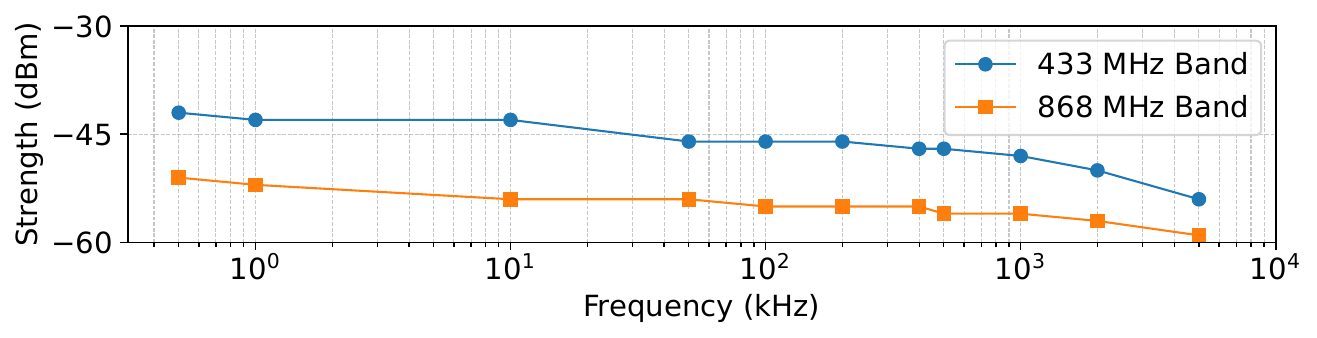}%
        \label{fig:ss_vs_bb_freq}
    }
    \vspace{-0.5mm}
    \subfloat[Baseband amplitude response]{%
        \includegraphics[width=0.95\linewidth]{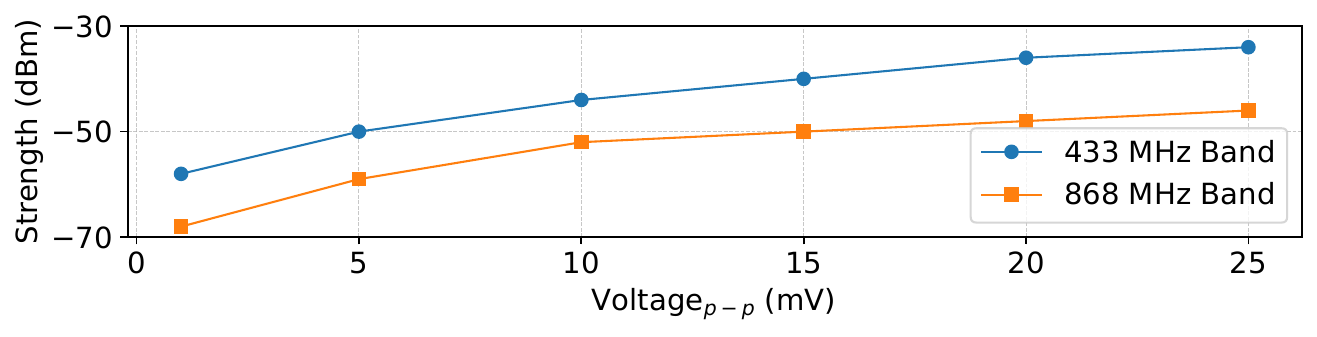}%
        \label{fig:ss_vs_bias_voltage}
    }
    \vspace{-1mm}
    \caption{Characterization of mixed signal strength. (a) The mixing bandwidth extends beyond \SI{1}{\mega\hertz} (b) The upconverted signal strength increases almost linearly with baseband amplitude.}
    \vspace{-4mm}
    \label{fig:ss_characterization}
\end{figure}

\subsection{Downconversion for Reception}
Complementary to transmission, \system functions as an autodyne mixer for reception. The diode's nonlinearity mixes incoming RF signals with the carrier to produce an intermediate frequency~(IF) signal, which is demodulated.

\fakepar{Harmonic autodyning} \system performs downconversion in two bands using a single device. In the \SI{433}{\mega\hertz} band, incoming signals near $F_{LO}$ mix with the fundamental to produce the IF signal. In the \SI{868}{\mega\hertz} band, incoming signals near $2F_{LO}$ mix with the internally generated harmonic, which acts as an effective LO despite no dedicated \SI{868}{\mega\hertz} oscillator present.

\begin{figure}[!t]
    \centering
    \includegraphics[width=0.95\linewidth]{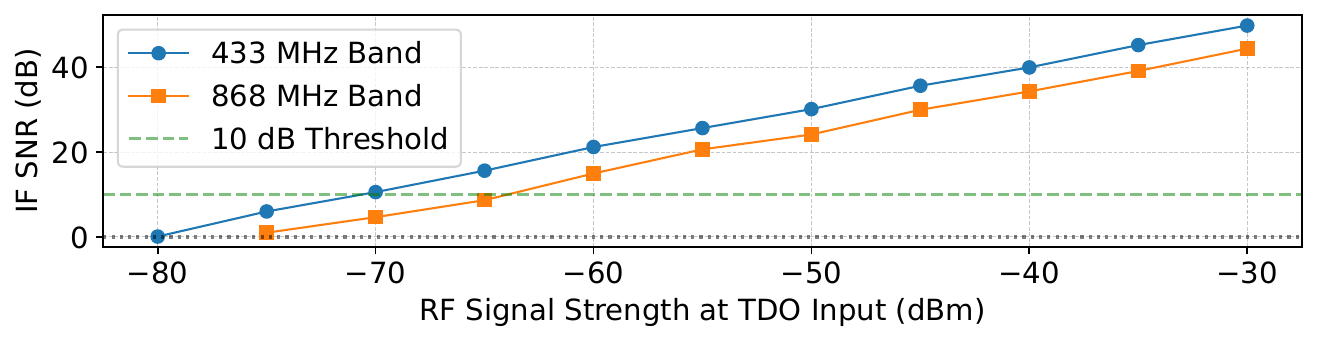}
    \vspace{-1mm}
    \caption{Receiver sensitivity. Using a standard \SI{10}{\decibel} SNR threshold, the system achieves a sensitivity of \SI{-70}{\decibel m} for the fundamental band and \SI{-62}{\decibel m} for the harmonic band.}
    \vspace{-5mm}
    \label{fig:sensitivity}
\end{figure}

\fakepar{Sensitivity} It refers to  the minimum signal strength that can be reliably detected and demodulated. We evaluate sensitivity using a cabled setup with two USRPs (one as the ED, the other as the transmitter), a Signal Hound BB60C spectrum analyzer, and a Mini-Circuits power combiner to interface with the \system frontend. We vary the transmitter  power and measure the strength of the downconverted signal, with the injection-locking signal at $-40$~dBm. The analyzer is configured with a resolution bandwidth (RBW) of \SI{100}{\hertz} and a span of \SI{20}{\kilo\hertz} to estimate the noise floor and calculate the signal-to-noise ratio~(SNR). Figure~\ref{fig:sensitivity} shows that the SNR of the downconverted signal increases with transmitter power.

For simple modulation schemes such as 2-FSK, an SNR of at least 10~dB is typically sufficient for low bit error rate (BER). Using this threshold, \system can reliably receive signals down to $-70$~dBm in the fundamental band. Due to reduced conversion efficiency at the harmonic, the sensitivity threshold for the \SI{868}{\mega\hertz} band is approximately $-62$~dBm. 

\begin{figure}[!t]
    \centering
    \includegraphics[width=0.95\linewidth]{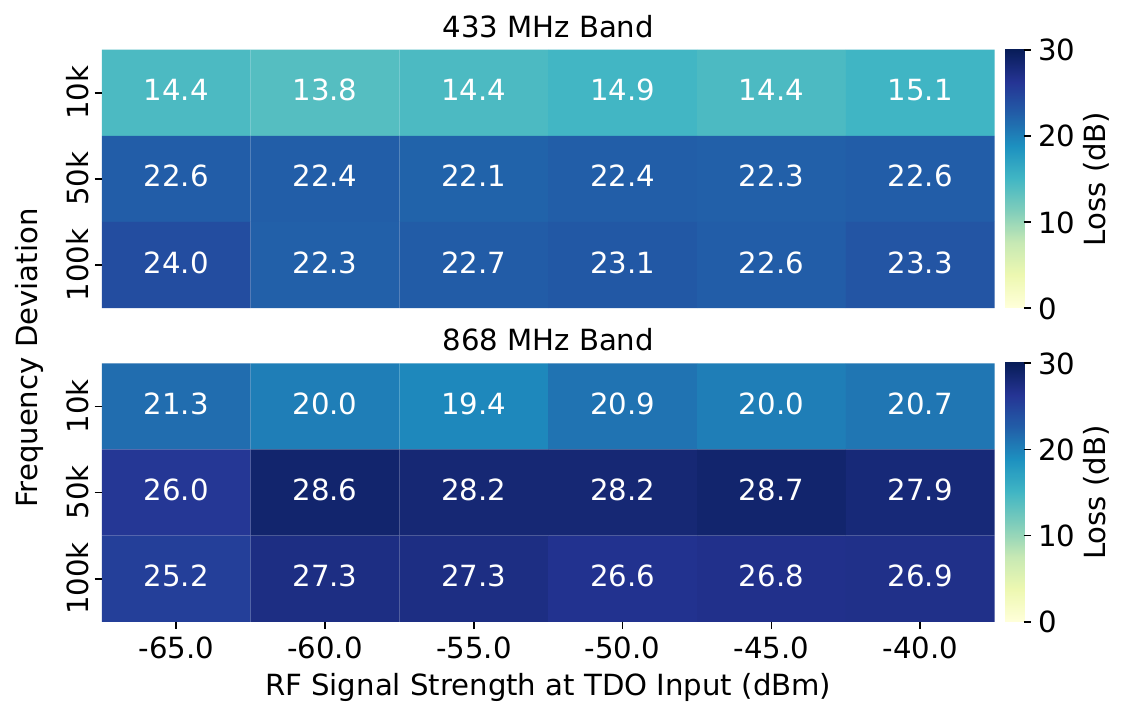}
    \vspace{-2mm}
    \caption{Conversion loss. The \SI{868}{\mega\hertz} harmonic band exhibits higher conversion loss (\SIrange{21}{28}{\decibel}) compared to the fundamental band (\SIrange{15}{23}{\decibel}) due to the weaker effective LO power at the frequency $2F_{LO}$.}
    \vspace{-6mm}
    \label{fig:conversion_loss}
\end{figure}

\fakepar{Conversion efficiency} To characterize dual-band performance, we measure conversion loss, defined as the ratio of RF input power to IF output power. Figure~\ref{fig:conversion_loss} compares conversion loss for both bands across different input powers and frequency deviations, using the same cabled setup as the previous experiment. As expected, the \SI{868}{\mega\hertz} band exhibits higher loss than the \SI{433}{\mega\hertz} band, confirming that mixing with the weaker harmonic ($2F_{LO}$) is less efficient than mixing with the fundamental ($F_{LO}$). Despite reduced conversion efficiency at the harmonic, the system maintains a sensitivity of $-62$~dBm without requiring the power or complexity of a secondary active mixer.



\section{Evaluation}

\fakepar{Injection-locking range} To determine the practical operating range of the \system frontend, we analyzed the maximum distance at which a ED can effectively injection-lock the tag. We observed that the locking threshold is \SI{-70}{\decibel m} at the TDO input; below this power level, the oscillator phase noise dominates, and the lock is lost. Prior works have reported similar thresholds~\cite{varshney2022judo,somix}. We used the ITU-R P.1411 urban line-of-sight model~\cite{iturP1411_13} to simulate the propagation loss in a typical deployment environment. The ED is assumed to transmit with an EIRP of \SI{10}{\decibel m}. Figure~\ref{fig:locking_range} compares the achievable locking range of a TDO tuned to a fundamental frequency of \SI{433}{\mega\hertz} versus a TDO tuned to \SI{868}{\mega\hertz}. The analysis demonstrates that at \SI{433}{\mega\hertz}, the tag receives sufficient power ($\ge \SI{-70}{\decibel m}$) to maintain locking up to a distance of \textbf{\SI{176}{\meter}}. In contrast, due to higher path loss at the higher frequency, an \SI{868}{\mega\hertz} oscillator would only sustain a locking range of \textbf{\SI{88}{\meter}}. This validates the \system architecture: by utilizing a lower fundamental frequency for synchronization, we effectively double the injection-locking range compared to an oscillator at higher fundamental frequency.

\begin{figure}[!t]
    \centering
    \includegraphics[width=0.95\linewidth]{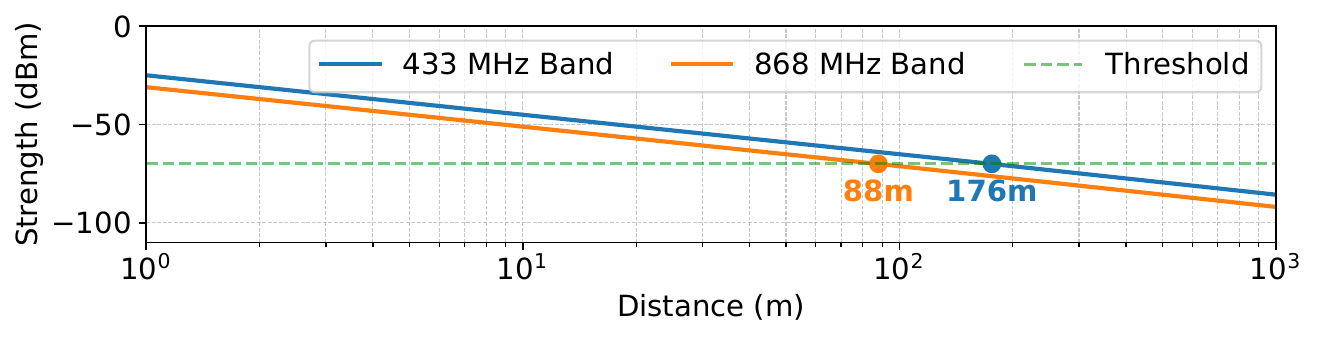}
    \vspace{-4mm}
    \caption{Injection locking the \system frontend at \SI{433}{\mega\hertz} extends the effective range to \SI{176}{\meter}, compared to only \SI{88}{\meter} for TDO tuned to \SI{868}{\mega\hertz} in an urban LoS deployment setting.}
    \vspace{-4mm}
    \label{fig:locking_range}
\end{figure}

\fakepar{Transmission range}
We evaluate the transmission range of \system by mixing the carriers generated by the TDO with a 2-FSK baseband waveform generated by an MSP430FR5969. The baseband is configured with a center frequency of \SI{90}{\kilo\hertz}, frequency deviation of \SI{10}{\kilo\hertz}, and data rate of \SI{3}{kbps}. It follows a fixed packet structure consisting of a \SI{4}{B}~(byte) preamble, a \SI{4}{B} synchronization word, a  \SI{1}{B} sequence number, a \SI{1}{B} node identifier, and a \SI{14}{B} payload. Cyclic redundancy check was disabled to enable analysis of corrupted packets. Experiments are conducted inside a university building in an indoor LoS setting with rich multipath due to walls and metallic structures. A USRP~B200 located \SI{10}{\meter} from the tag serves as the ED for injection locking. We use two TI~CC1310 receivers to simultaneously receive the transmissions at the fundamental and harmonic bands while varying the receiver distance from \SIrange{1}{30}{\meter}. We use Abracon \SIrange{433}{960}{\mega\hertz} whip antennas having a gain of \SI{3}{dBi} for the \system transmitter, ED and receivers as shown in Figure~\ref{fig:fig:tx_setup}. Figure~\ref{fig:rssi_ber} shows the measured RSSI and BER averaged over 1000 packets across three trials. \system maintains BER $\le 10^{-2}$ up to \SI{30}{\meter} in both bands. As expected, the \SI{868}{\mega\hertz} link exhibits lower RSSI than \SI{433}{\mega\hertz} due to the weaker harmonic component, but remains decodable. This also highlights the reliability of \system to receive at the harmonic frequency, in case the fundamental undergoes frequency-selective fading. We did not evaluate beyond \SI{30}{\meter} due to space constraints.

\begin{figure}[!t]
    \centering
    \subfloat[Experimental setup]{%
        \includegraphics[width=0.9\linewidth]{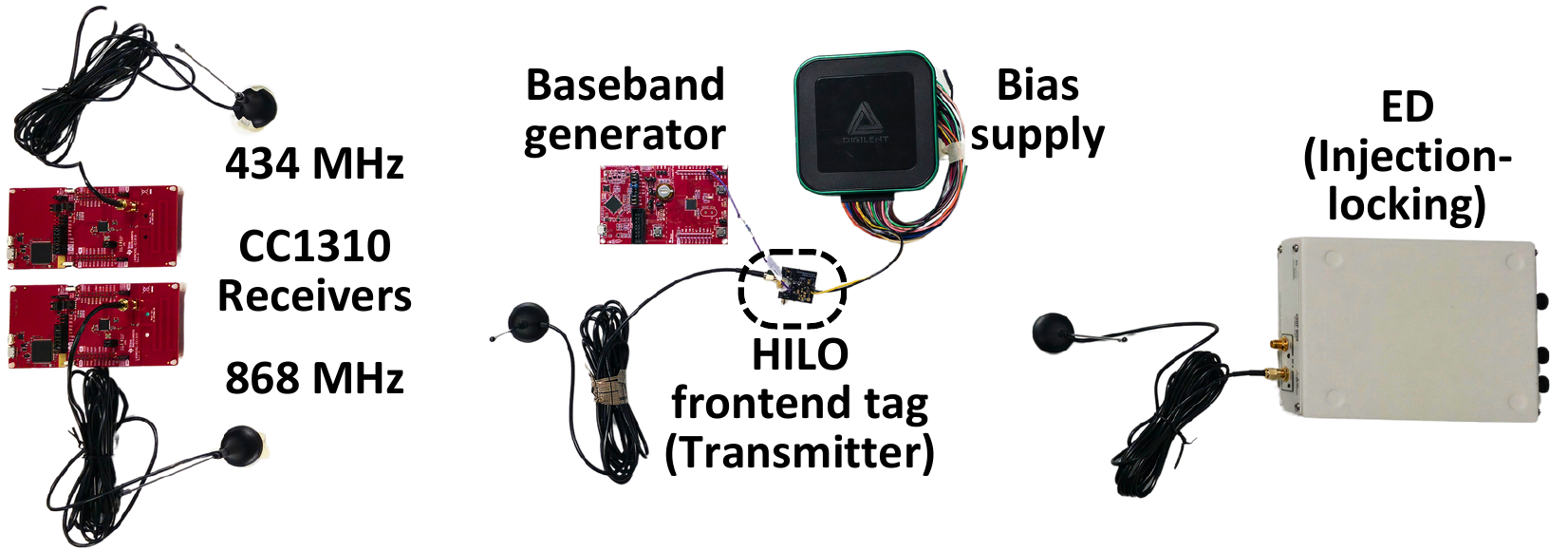}%
        \label{fig:fig:tx_setup}
    }
    \vspace{-0.5mm}
    \subfloat[Transmission range]{%
        \includegraphics[width=0.95\linewidth]{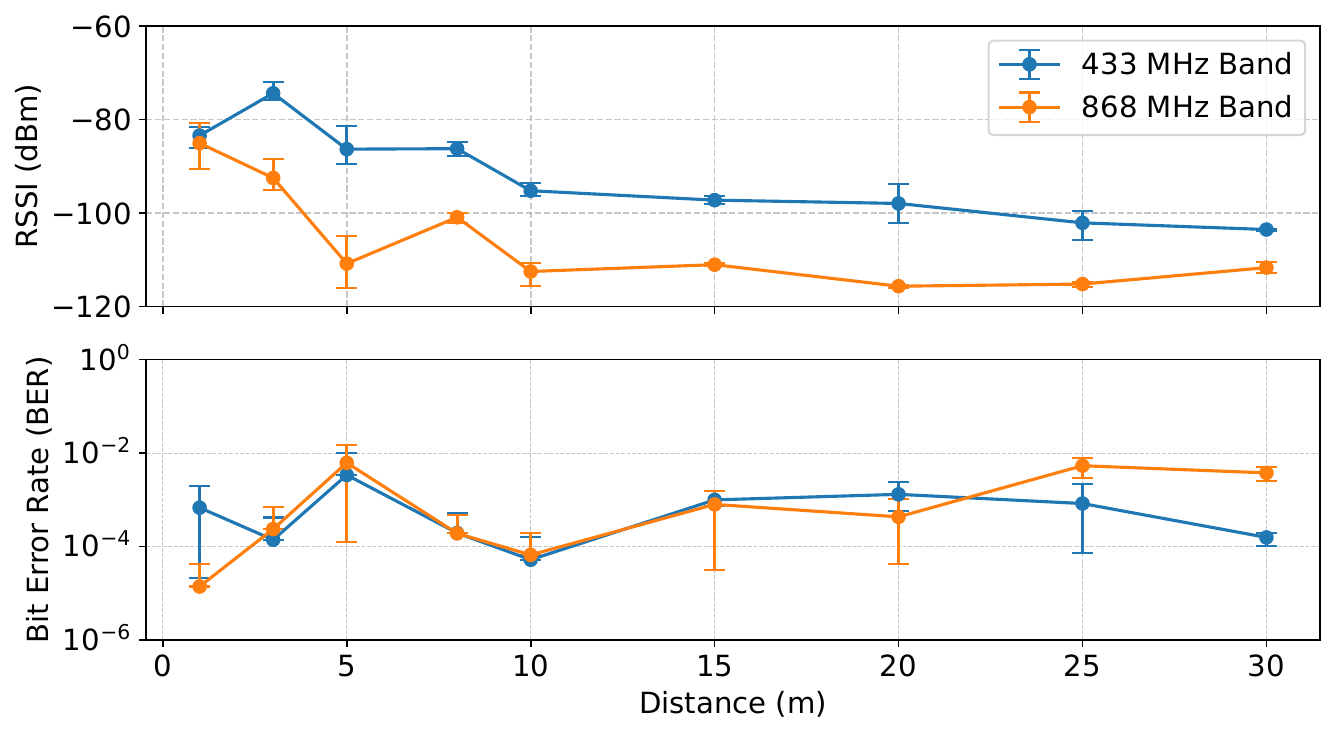}%
        \label{fig:rssi_ber}
    }
    \vspace{-1mm}
    \caption{We conduct experiments in complex indoor line-of-sight environment to evaluate the transmission range. The harmonic link is consistently weaker yet remains decodable up to \SI{30}{\meter}. The non-monotonic trends highlight frequency-selective fading and the two carriers can be impaired differently.}
    \vspace{-6mm}
    \label{fig:tx_range}
\end{figure}

\fakepar{Power consumption} We measured the power consumption using the Keysight B2902A precision source/measure unit. The \system frontend is biased at a supply voltage of \SI{350}{\milli\volt} and draws a measured current of \SI{0.6}{\milli\ampere}, resulting in a static power consumption of \SI{210}{\micro\watt}. The power budget of the complete system would additionally include the baseband generator for transmission~($\approx$\SI{100}{\micro\watt}) and the demodulator pipeline for reception.
Importantly, this \SI{210}{\micro\watt} power budget supports simultaneous dual-band transmission or reception. Unlike conventional multi-band architectures that require a second RF chain to toggle between bands, \system leverages the oscillator's inherent nonlinearities to generate the \SI{868}{\mega\hertz} carrier alongside the \SI{433}{\mega\hertz} fundamental without additional active circuitry. This effectively amortizes the energy cost across two frequency bands and enhances energy efficiency.

\section{Discussion and Future Work}

\fakepar{Interference mitigation} Figure~\ref{fig:harmonics_spectrum} shows higher-order harmonics arising from the tunnel diode's nonlinearity. While the second harmonic ($2F_{LO}$) is intentionally exploited for communication, higher-order components (e.g., $3F_{LO}$ and $4F_{LO}$) appear approximately 25~dB and 35~dB below the fundamental. These constitute unintentional emissions that must be suppressed to meet regulatory spectral masks in commercial deployments. Our time-domain measurements in Figure~\ref{fig:timedomain} show that the oscilloscope's \SI{1}{\giga\hertz} bandwidth naturally attenuates these components, suggesting that a simple low-pass filter or diplexer can provide sufficient suppression. A diplexer additionally enables separation of the fundamental and harmonic carriers, allowing band-optimized antennas for improved efficiency. We leave this as our future effort. 

\begin{figure}[!t]
    \centering
    \includegraphics[width=\linewidth]{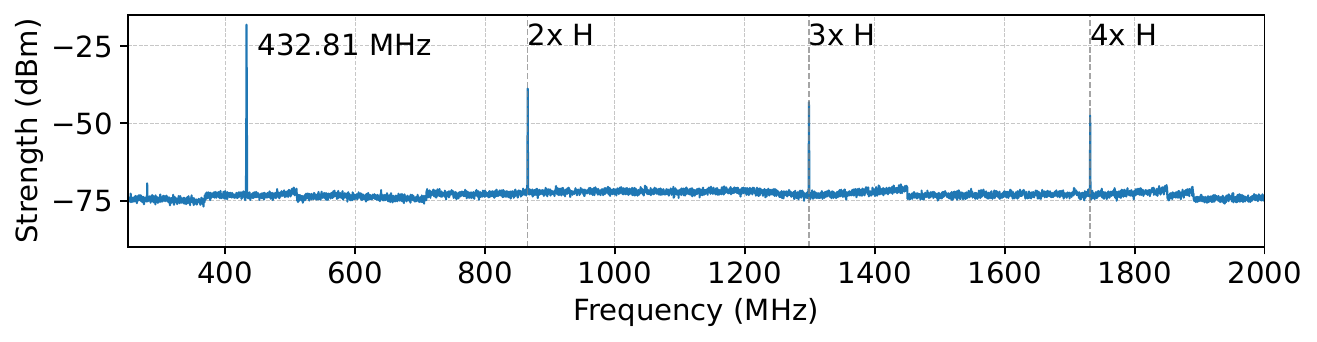}
    \vspace{-8mm}
    \caption{Wideband spectrum of the TDO. While the second harmonic ($2F_{LO}$) is exploited for communication, higher-order components are present and require filtering for regulatory compliance.}
    \vspace{-6mm}
    \label{fig:harmonics_spectrum}
\end{figure}

\fakepar{Supporting dual-band reception}
While we demonstrate dual-band downconversion at both $F_{LO}$ and $2F_{LO}$, this work does not implement a complete end-to-end receiver. A full receiver would additionally require analog baseband amplification and filtering, ADC sampling, and digital demodulation to decode packets. To fully exploit frequency diversity, future systems could incorporate receive-side combining strategies (e.g., selection or maximal-ratio combining) that account for unequal SNR and band-specific impairments. Exploring these remains an important direction for future work.

\fakepar{Eliminating emitter-device} \system currently relies on injection locking from an ED for TDO stability, thus partially retaining the delegation architecture common to low-power communication systems. Recent work has taken significant steps toward improving TDO stability and eliminating the need for an ED. AudioCast~\cite{audiocast} combines a number of design choices, including purposefully operating at a lower frequency band, voltage filtering, and RF shielding, to stabilize the TDO at FM-band frequencies. A subsequent effort, SharpPeak~\cite{sharppeak_sensys26}, argues that controlled voltage transitions lead to improved TDO stability. However, all these systems rely on LC networks using discrete off-the-shelf components, which limits the Q-factor and results in long-term frequency instability beyond what commodity wireless standard requires, particularly under changing temperature and bias voltage conditions. Our recent work, M\textsuperscript{2}~\cite{m2_mobisys26}, couples a tunnel diode with a high-Q SAW resonator, achieving orders of magnitude better stability than SharpPeak and AudioCast. M\textsuperscript{2} also generates harmonics while retaining the SoM and autodyning properties, making it a promising foundation for designing standalone dual-band transceivers, which we leave as a future effort.

\section{Conclusion}

We present \systemnospace, a tunnel-diode-based RF frontend that enables dual-band operation using a single injection-locked oscillator. By locking the fundamental near \SI{433}{\mega\hertz}, \system stabilizes and exploits its naturally generated second harmonic near \SI{868}{\mega\hertz}, eliminating the need for a second RF chain or frequency synthesis. A single baseband input is self-mixed onto both carriers for concurrent dual-band transmission, and the same frontend acts as an autodyne mixer to downconvert signals in both bands. This frequency-diverse operation improves robustness to frequency-selective fading and interference. Our prototype achieves reliable dual-band transmission beyond \SI{30}{m} in LoS while consuming under \SI{210}{\micro\watt} of power.

\section*{Acknowledgements}
We thank the anonymous reviewers for their insightful comments. This work is funded by a Tier-1 grant from the Ministry of Education (A-8001661-00-00), a startup grant from ODPRT (A-8000277-00-00), and an unrestricted gift from Google through their Research Scholar Program (A-8002307-00-00-00), hosted at the National University of Singapore.

\bibliographystyle{IEEEtran}
\bibliography{references}

\end{document}